\documentclass[prb,twocolumn]{revtex4-1}
\usepackage{amsmath,amsfonts}    
\usepackage{graphicx}   
\usepackage{color}
\usepackage{hyperref}  
\usepackage{epstopdf}
\usepackage{subfig}
\DeclareGraphicsExtensions{.pdf,.jpeg,.png}

\providecommand{\ds}{\displaystyle}
\providecommand{\w}{\ensuremath{\omega} }

\providecommand{\bo}[1]{\ensuremath{\mathcal{O}\left(#1\right)}}
\providecommand{\grad}{\ensuremath{\nabla}}
\providecommand{\sech}{\text{sech}}
\providecommand{\abs}[1]{\ensuremath{\left| #1 \right|}}

\providecommand{\lap}{\ensuremath{\nabla^2}}

\begin{document}
\title{Analytical theory of modulated magnetic solitons}
\author{L. D. Bookman}
 \email{ldbookma@ncsu.edu}
\author{M. A. Hoefer}
\email{mahoefer@ncsu.edu}
\affiliation{Department of Mathematics, North Carolina State University, Raleigh, North Carolina 27695, USA}
\begin{abstract}
  Droplet solitons are coherently precessing solitary waves that have
  been recently realized in thin ferromagnets with perpendicular
  anisotropy.  In the strongly nonlinear regime, droplets can be well
  approximated by a slowly precessing, circular domain wall with a
  hyperbolic tangent form.  Utilizing this representation, this work
  develops a general droplet modulation theory and applies it to study
  the long range effects of the magnetostatic field and a nanocontact
  spin torque oscillator (NC-STO) where spin polarized currents act as
  a gain source to counteract magnetic damping.  An analysis of the
  dynamical equations for the droplet's center, frequency, and phase
  demonstrates a negative precessional frequency shift due to long
  range dipolar interactions, dependent on film thickness.  Further
  analysis also demonstrates the onset of a saddle-node bifurcation at
  the minimum sustaining current for the NC-STO.  The basin of
  attraction associated with the stable node demonstrates that spin
  torque enacts a restoring force to excursions of the droplet from
  the nanocontact center, observed previously in numerical
  simulations.  Large excursions lead to the droplet's eventual decay
  into spin waves.
\end{abstract}
\maketitle

\section{Introduction}
\label{sec:introduction}

Magnetic materials are a rich setting for the study of nonlinear,
coherent structures.  Previously, magnetic excitation mechanisms were
predominantly limited to the application of magnetic fields.
Nowadays, spin polarized currents
\cite{slonczewski_excitation_1999,berger_emission_1996} are commonly
used to create, manipulate, and control nano-scale magnetic
excitations such as domain walls \cite{boulle_current-induced_2011}
and vortices \cite{petit-watelot_commensurability_2012}, promising
candidates for applications
\cite{allwood_magnetic_2005,pribiag_magnetic_2007}.  Recently, a
strongly nonlinear, coherently precessing localized mode termed a
droplet soliton was observed in a nanocontact spin torque oscillator
(NC-STO) with perpendicular magnetic anisotropy
\cite{mohseni2013spin}.  The observed droplet exhibited a number of
intriguing features including sub-ferromagnetic resonance frequencies,
low frequency modulations, and an almost complete reversal of the
magnetization within its core, hence a perimeter precession angle of
180 degrees.  Supporting micromagnetic simulations demonstrated that
the low frequency modulations could be due to an oscillation of the
droplet within the nanocontact resulting from some restoring force.
Previous theoretical studies of the NC-STO droplet neglected long
range dipolar interactions and observed a drift instability whereby
the droplet was ejected from the nanocontact \cite{hoefer2010theory}.
Motivated by these observations, we analytically and numerically study
general modulations of a large amplitude magnetic droplet soliton's
precessional frequency, phase, and position with particular emphasis
on the effects due to magnetostatics, magnetic damping, and spin
torque in the nanocontact geometry.

Basic properties of unperturbed, conservative droplets have been
extensively studied \cite{kosevich1990magnetic}.  A central assumption
is that of symmetry which gives rise to conserved quantities and a
family of soliton solutions.  A stationary droplet soliton can be
parameterized by its center, initial phase and precession frequency.
In physical problems of interest, these high symmetry, idealized
conditions are typically not met. Nevertheless, due to their robust
qualities, localized structures may persist.  In the context of weak,
symmetry breaking perturbations, a modulation theory can be developed
whereby the soliton's parameters are allowed to vary adiabatically in time
\cite{kivshar_dynamics_1989}.  The resulting soliton modulation
equations are analogous to Thiele's equation \cite{thiele_steady_1973}
for the motion of a magnetic domain or vortex.  One symmetry breaking
example is the dissipative droplet soliton, excited in the
spin-transfer torque (STT) driven NC-STO where translational
invariance and time reversal symmetry are broken
\cite{hoefer2010theory}.  This is precisely the soliton observed in
the previously discussed experiment \cite{mohseni2013spin}.  % Symmetry
% breaking in this case reduces the size of the soliton solution
% family.
In steady state, the balance between STT forcing and damping
centers the droplet within the nanocontact and selects a specific
frequency but the phase is still arbitrary. Modulation theory
generalizes the steady state conditions by allowing for slow temporal
variations of the conservative droplet's parameters due to the
symmetry breaking perturbations.  In effect, the droplet is treated as
a slowly moving, precessing dipole particle.

In order to make analytical progress tractable and to enable efficient
micromagnetic simulations, magnetic soliton studies often neglect the
long range component of the magnetostatic field, inherent in any
magnetic sample that exhibits a nonuniform magnetization distribution.
This is a reasonable approximation in the case of very thin, extended
magnetic films \cite{gioia_micromagnetics_1997} where the
magnetostatic field takes the local form $-M_z \mathbf{z}$ ($M_z$ is
the component of the magnetization perpendicular to the film);
however, thickness-dependent, long range corrections can be important
\cite{garcia-cervera_one-dimensional_2004}.  These corrections lead to
a breaking of phase invariance \cite{papanicolaou_dynamics_1991}.

Here, we present the stationary droplet soliton modulation equations
for symmetry breaking perturbations.  An approximate, analytical
representation of the conservative droplet in the strongly nonlinear
regime is discussed and used to greatly simplify the modulation
equations.  The effects of the long range magnetostatic field, a
NC-STO, and magnetic damping are studied in detail by a dynamical
systems analysis of the modulation equations.  The long range
component of the magnetostatic field is shown to give rise to a
thickness-dependent, negative precessional frequency shift of the
droplet.  These dynamics are independent of and do not alter the
effects due to the NC-STO and damping.  The dissipative droplet
soliton is identified as the stable node of a saddle-node bifurcation
for sufficiently large spin torque.  Consequently, spin torque
provides a restoring force to deviations in droplet frequency and
position from the nanocontact center.  This analytical prediction
explains the micromagnetic observation of a restoring force and the
corresponding slow droplet modulations observed in
Ref.~\onlinecite{mohseni2013spin}.  However, large deviations can lead
to decay to small amplitude spin waves, helping to explain the
previously observed droplet drift instability \cite{hoefer2010theory}.

\section{Model Equation and Nondimensionalization}
\label{sec:model-equation}

The mathematical model considered here is the following torque
equation for the vector field magnetization $\mathbf{M}$
\begin{equation}
   \label{eq:LL}
  \begin{split}
    \frac{\partial \mathbf{M}}{\partial t} &= - \abs{\gamma} \mu_0  \mathbf{M}\times
    \mathbf{H}_{\rm eff}  + \mathbf{P}, \\
    \mathbf{H}_{\rm eff} & = \frac{2 A }{\mu_0 M_{\mathrm{s}}^2} \lap \mathbf{M} +
    \left(H_0 + \frac{2 K_{\mathrm{u}}}{\mu_0 M_{\mathrm{s}}^2} M_z\right) \mathbf{z} +
    \mathbf{H}_{\rm{m}}.
  \end{split}
\end{equation}
The ferromagnetic material is taken to be of infinite extent in the
$x$-$y$ directions and of finite thickness $\delta$ in $z$.  The
parameters are the gyromagnetic ratio $\gamma$, the permeability of
free space $\mu_0$, the exchange stiffness parameter $A$, the
perpendicular magnetic field amplitude $H_0$, the crystalline
anisotropy constant $K_{\mathrm{u}}$, and the saturation magnetization
$M_{\mathrm{s}}$.  $\mathbf{P}$ represents any perturbation that
maintains the magnetization's total length, i.e. $\mathbf{P}\cdot
\mathbf{M} \equiv 0$.  The boundary conditions are $\lim_{x^2+y^2 \to
  \infty} \mathbf{M} = M_{\mathrm{s}} \mathbf{z}$ and $\partial
\mathbf{M}/\partial z = 0$ when $z = \pm \delta/2$.
$\mathbf{H}_{\rm{m}}$ is the magnetostatic field resulting from
Maxwell's equations.   

As derived in Ref. \onlinecite{garcia-cervera_one-dimensional_2004}, the
magnetostatic energy for a $z$ independent magnetization can be given
in Fourier space as
\begin{equation}
  \label{eq:1}
  \begin{split}
    \mathcal{E}_{\mathrm{m}} = \frac{\delta}{2}
    \int_{\mathbb{R}^2} \Bigg \{ &\frac{|\mathbf{k} \cdot
      \widehat{\mathbf{M}}_\perp |^2}{k^2}[1 -
    \widehat{\Gamma}(k\delta)] \\
    &+ |\widehat{M_z - M_{\mathrm{s}}}|^2
    \widehat{\Gamma}(k \delta)
    \Bigg \} \, \mathrm{d} \mathbf{k},
  \end{split}
\end{equation}
where
\begin{equation}
  \label{eq:2}
  \widehat{\Gamma}(\kappa) = \frac{1 - e^{-\kappa}}{\kappa} .
\end{equation}
Computing the negative variational derivative of
$\mathcal{E}_{\mathrm{m}}$ with respect to $\mathbf{M}$ and expanding
$\widehat{\Gamma}(k\delta)$ for $|k\delta| \ll 1$ yields
 the two-dimensional (2D), film thickness averaged
magnetostatic asymptotic approximation
\begin{equation}
  \label{eq:1}
  \begin{split}
    \mathbf{H}_{\mathrm{m}} &\sim -M_z \mathbf{z} + \frac{\delta}{2}
    \mathbf{H}_{\mathrm{nl}}, \\
    \mathbf{H}_{\mathrm{nl}} &= \mathbf{z} \sqrt{-\nabla^2} (M_z -
    M_{\mathrm{s}}) + \frac{1}{\sqrt{-\nabla^2}}
    \nabla (\nabla \cdot \mathbf{M}_\perp).
  \end{split}
\end{equation}
The magnetostatic field is composed of the usual local term $-M_z
\mathbf{z}$ and a long range, nonlocal contribution $\frac{\delta}{2}
\mathbf{H}_{\mathrm{nl}}$.  We define $\mathbf{M}_\perp = (M_x,M_y)$
and assume $\delta$ to be small relative to the typical transverse
wavelength of excitation, i.e., the exchange length $l_{\mathrm{ex}} =
\sqrt{2 A/(\mu M_{\mathrm{s}}^2)}$.  The operators are interpreted in
Fourier space, e.g., $\widehat{\sqrt{-\nabla^2} f} = |\mathbf{k}|
\widehat{f}$ and $\widehat{f}(\mathbf{k})$ is the two-dimensional
Fourier transform of $f$ at wavevector
$\mathbf{k}$.  % We will absorb the long range
% magnetostatic correction $- \delta \mathbf{M}\times
% \mathbf{H}_{\mathrm{nl}}/2$ into the small perturbation $\mathbf{P}$.
Long range magnetostatic corrections have also been used to study
domain patterns and vortices in materials with easy-plane anisotropy
\cite{garcia-cervera_effective_2001}.  In order to nondimensionalize
the equation, we introduce the dimensionless quality factor $Q= 2
K_{\mathrm{u}}/(\mu_0 M_{\mathrm{s}}^2)$ that measures the strength of
the uniaxial, crystalline anisotropy.  The quality factor is assumed
to be greater than unity in order to guarantee the existence of
droplet solutions in the unperturbed ($\mathbf{P}=0$, $\delta = 0$)
problem \cite{kosevich1990magnetic}. Nondimensionalizing time by
$[\abs{\gamma}\mu_0 M_{\mathrm{s}} ( Q - 1)]^{-1}$, lengths by
$l_{\mathrm{ex}}/\sqrt{Q-1}$, fields by $M_{\mathrm{s}}(Q-1)$, and
setting $\mathbf{m} = \mathbf{M}/M_{\mathrm{s}}$, eq.~\eqref{eq:LL}
becomes the 2D model
\begin{equation}
  \label{eq:nondimLL}
  \begin{split}
    \frac{\partial \mathbf{m}}{\partial t} &= - \mathbf{m} \times
    \left( \lap \mathbf{m} + (m_z +
      h_0) \mathbf{z}\right) + \mathbf{p}, \\
    \mathbf{p} &= \frac{\mathbf{P}}{|\gamma| \mu_0
      M_{\mathrm{s}}^2(Q-1)} - \frac{\delta}{2} \mathbf{m} \times
    \mathbf{h}_{\mathrm{nl}}, \quad (x,y) \in \mathbb{R}^2 .
  \end{split}
\end{equation}
Small amplitude, spin wave excitations to the uniform state
$\mathbf{m} = \mathbf{z}$ of the unperturbed problem $\mathbf{p}
= 0$ admit the exchange dispersion relation
$\omega(\mathbf{k}) = 1 + k^2$ where $\omega(0) = 1$ represents the
scaled ferromagnetic resonance frequency.  The rest of this work
concerns soliton dynamics associated with eq.~\eqref{eq:nondimLL}.

\section{Approximate Droplet Solution}
\label{sec:appr-dropl-solut}

First we consider droplet soliton solutions of eq.~\eqref{eq:nondimLL}
when $\mathbf{p} = 0$ representing an infinitely thin, undamped
ferromagnet with strong perpendicular, uniaxial anisotropy. It is
convenient to represent $\mathbf{m}$ in spherical coordinates by the
radial unit vector, $\mathbf{m} = [ \cos \Phi \sin\Theta,
\sin \Phi \sin \Theta, \cos\Theta]$.  In these coordinates, the
stationary droplet soliton solution to eq.~\eqref{eq:nondimLL} is the
positive, monotonically decaying solution of the boundary value
problem
\begin{equation}
  \label{eq:droplet} 
  \begin{cases}
    \ds- \left(\frac{d^2}{d \rho^2} + \frac{1}{\rho} \frac{d}{d\rho}
    \right) \Theta_0 + \sin \Theta_0 \cos \Theta_0 - \omega \sin
    \Theta_0 = 0, \vspace{2mm}\\
    \ds \frac{d \Theta_0 }{d \rho} (\mathbf{x}_0; \omega) = 0,
    \hspace{1.79cm} \lim_{\rho \rightarrow \infty} \Theta_0( \rho;
    \omega) = 0,
  \end{cases}
\end{equation}
where $\Theta = \Theta_0$, $\Phi = (\omega + h_0) t + \Phi_0$, $\rho$
is the radial distance from $\mathbf{x}_0$, and $0 < \omega < 1$
\cite{kosevich1990magnetic}.  Consequently, the stationary droplet is
parameterized by frequency $\omega$, the initial phase $\Phi_0$, and
the initial droplet center coordinates $\mathbf{x}_0$, the latter
generated by invariances with respect to azimuthal rotations of
$\mathbf{m}$ and translations.  Droplets can also be made to
propagate\cite{hoefer_propagating_2012}.  While this work is concerned
with stationary droplets, we will use the propagating solution in the
Appendix to implement the modulation theory.  The boundary conditions
in \eqref{eq:droplet} arise from the far field decay condition applied
to \eqref{eq:LL} and the requirement that \eqref{eq:droplet} remain
finite near the droplet center.

\begin{figure}
  \centering
  \includegraphics[width=\columnwidth]{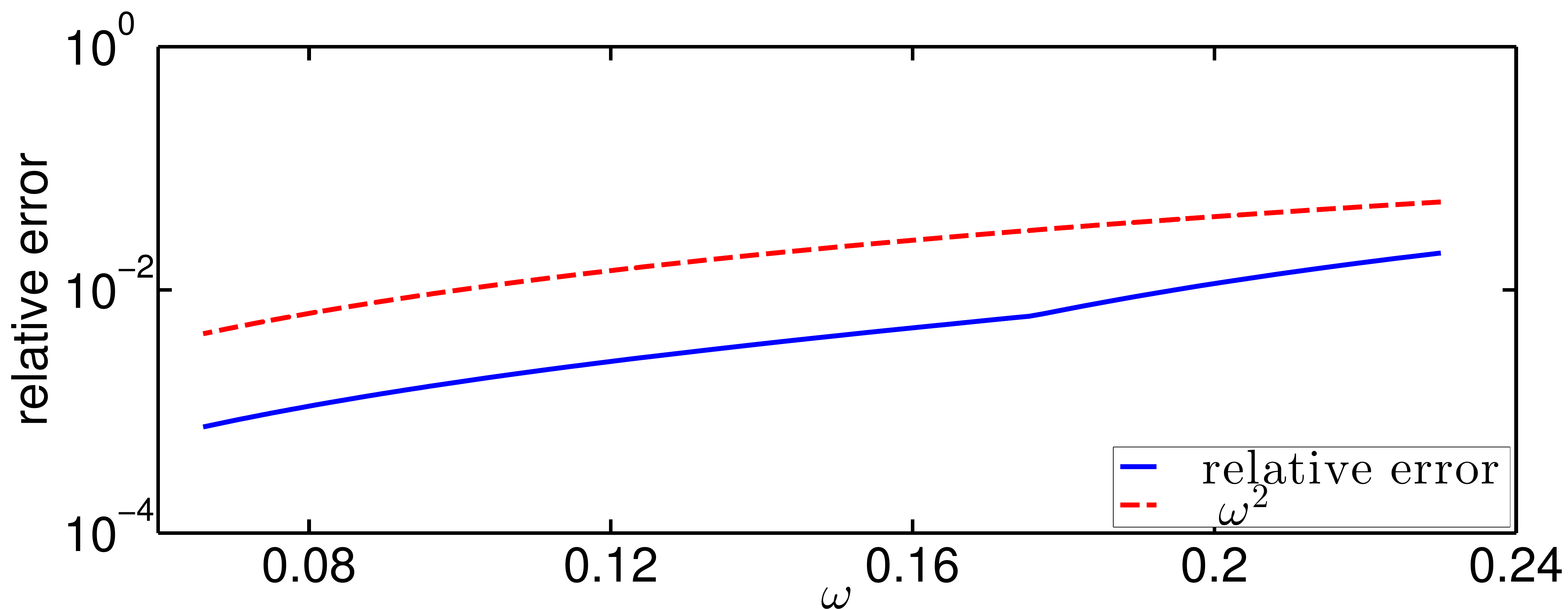}
  \caption{Relative difference between approximate and numerically
    computed droplets (solid).  $\omega^2$ (dashed) for comparison of
    convergence order.}
  \label{fig:numerical_validation}
\end{figure}

In the small $\omega$ regime, the droplet profile takes the
approximate form \cite{kosevich__1986,ivanov__1989}
\begin{equation}
  \cos \Theta_0 = \tanh (\rho - 1/\omega), \quad 0 < \omega \ll
  1. \label{eq:dropletsol}   
\end{equation}
We have derived this approximate solution using singular perturbation
theory (Appendix~\ref{sec:approx-droplet}) and find it to be accurate to
$\mathcal{O}(\omega^2)$ and uniformly valid for all $\rho \in
(0,\infty)$.  The error in the approximate solution
\eqref{eq:dropletsol} is shown in
Fig.~\ref{fig:numerical_validation}.
Based on the form of the solution \eqref{eq:dropletsol}, the small
$\omega$ droplet takes the form of a slowly precessing (absent the
applied field), circular domain wall with radius $1/\omega$.
Expanding this approximate solution for the droplet around $\rho =0$,
we observe $\Theta_0(0) = \pi$ to all orders in $\omega$.  However,
the magnetization at the center of the small $\omega$ droplet 
cannot equal $-\mathbf{z}$, due to its nontopological structure
\cite{kosevich1990magnetic}.  For the rest of this work, we will
consider the small $\omega$ regime and use the approximate form
\eqref{eq:dropletsol} for the droplet.

\section{Modulation Equations}
\label{sec:modulation-equations}

We now consider the effects of small symmetry breaking perturbations
$\mathbf{p}$ in eq.~\eqref{eq:nondimLL} on the droplet
\eqref{eq:dropletsol}.  The perturbation has spherical components
$p_\Theta = \mathbf{p} \cdot \mathbf{\Theta}$ and $p_\Phi = \mathbf{p}
\cdot \mathbf{\Phi}$.  The principle effects can be captured by
allowing the droplet's parameters to vary adiabatically in time, e.g.,
$\omega = \omega(t)$ with $|d\omega/dt| \ll 1$.  The method of
multiple scales allows for the determination of their evolution.
Following the approach in Ref.~\onlinecite{weinstein1985modulational}
developed for perturbations to a Nonlinear Schr\"{o}dinger soliton, we
linearize eq.~\eqref{eq:nondimLL} around a droplet and apply
solvability conditions at $\mathcal{O}(|\mathbf{p}|)$ to determine the
modulation equations (Appendix~\ref{sec:mod-eqns}),
\begin{align}
  \frac{d\omega}{d t} &= - \frac{\omega^3}{4 \pi} \int_{\mathbb{R}^2}
  \sech (\rho - 1/\omega) p_\Theta
  \, \mathrm{d}\mathbf{x}, \label{eq:mod_w_smallw}\\
  0& = \int_{\mathbb{R}^2} \sech(\rho - 1/\omega)
  p_\Phi \begin{pmatrix} \cos \varphi \\ \sin \varphi \end{pmatrix}
  \, \mathrm{d}\mathbf{x},\label{eq:mod_c_smallw} \\
  \frac{d \Phi_0}{d t}& =\frac{\omega}{4 \pi} \int_{\mathbb{R}^2}
  \sech(\rho - 1/\omega) p_\Phi \, \mathrm{d}
  \mathbf{x},\label{eq:mod_phi_smallw}\\
   \frac{d\mathbf{x}_0}{d t}& = \frac{\omega}{2 \pi} \int_{\mathbb{R}^2}
  \sech(\rho - 1/\omega) p_\Theta \begin{pmatrix}
    \cos \varphi \\ \sin \varphi \end{pmatrix} \,\mathrm{d}
  \mathbf{x}, \label{eq:mod_x_smallw}
\end{align}
where the perturbation $(p_\Theta,p_\Phi)$ is evaluated at the droplet
solution \eqref{eq:dropletsol} and $(\rho,\varphi)$ are the polar
coordinates for the domain $\mathbb{R}^2$.  Note that
eq.~\eqref{eq:mod_c_smallw} is not an evolution equation, but rather
serves as a constraint on admissible perturbations. When this
constraint is not satisfied, a nonzero momentum can be generated and
the droplet no longer remains stationary.  The stationary assumption
is essential not only to the modulation equations themselves but also to
the small $\omega$ approximation given in eq.~\eqref{eq:dropletsol}
and therefore a different set of modulation equations is required for
the non-stationary case.  For example, a magnetic field gradient will
accelerate a stationary droplet \cite{hoefer_propagation_2012}.

\subsection{Long Range Magnetostatic Perturbation}
\label{sec:long-range-magn}

\begin{figure}
  \centering
  \includegraphics[width=\columnwidth]{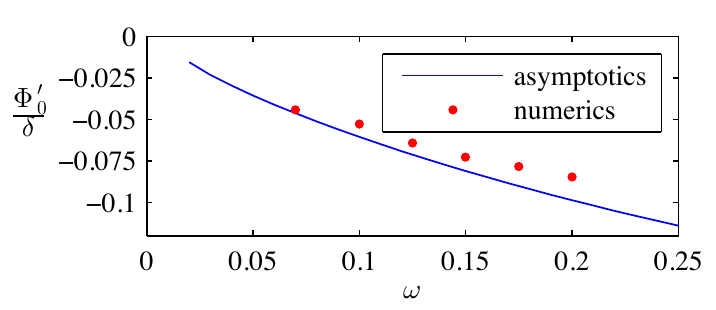}
  \caption{Negative frequency shift due to long range, thickness
    dependent magnetostatic corrections.  Equation \eqref{eq:3}
    (solid) and micromagnetic simulations with $\delta = 0.1$
    (dots).}
  \label{fig:freq_shift}
\end{figure}

We now investigate specific perturbations of physical relevance.
First, the long range magnetostatic field is considered.  After
applying modulation theory, we find that the contribution to
eqs.~\eqref{eq:mod_w_smallw}-\eqref{eq:mod_x_smallw} takes the form
\begin{equation}
  \label{eq:2}
  p_\Theta = 0, \quad p_\Phi = - \delta \sin \Theta_0
  \sqrt{-\nabla^2}(1 - \cos \Theta_0)/2 .
\end{equation}
Consequently, thickness dependent magnetostatic effects only enter in
eqs.~\eqref{eq:mod_c_smallw} and \eqref{eq:mod_phi_smallw}.  The
constraint equation \eqref{eq:mod_c_smallw} is automatically satisfied
because $p_\Phi$ depends only on $\rho$ so the $\varphi$ integrals
vanish.  What is left is the expression for the slowly varying phase
$\Phi_0$.  Evaluating \eqref{eq:mod_phi_smallw} with \eqref{eq:2}
represents a precessional frequency shift of the droplet
\begin{equation}
  \label{eq:3}
  \begin{split}
    \frac{d \Phi_0}{dt} = -\frac{\delta \omega}{4} &\int_0^\infty \mathrm{sech}^2(\rho -
    1/\omega) \\
    &\quad \times \{ \sqrt{-\nabla^2} [ 1 - \tanh(\rho-1/\omega) ] \} \rho \,
    \mathrm{d} \rho .
  \end{split}
\end{equation}
The total droplet frequency is $h_0 + \omega + \Phi_0'$.  Since the
integrand is strictly positive for $\rho \in (0,\infty)$,
eq.~\eqref{eq:3} represents a negative frequency shift which is
plotted in Fig.~\ref{fig:freq_shift} as a function of $\omega$.
Micromagnetic simulations (Appendix~\ref{sec:num-methods}) yield good, asymptotic
$\mathcal{O}(\delta \omega)$ agreement as expected.

\subsection{NC-STO and Damping Perturbations}
\label{sec:nc-sto-damping}

\begin{figure}
  \centering
  \subfloat{\includegraphics[width=.245\textwidth]{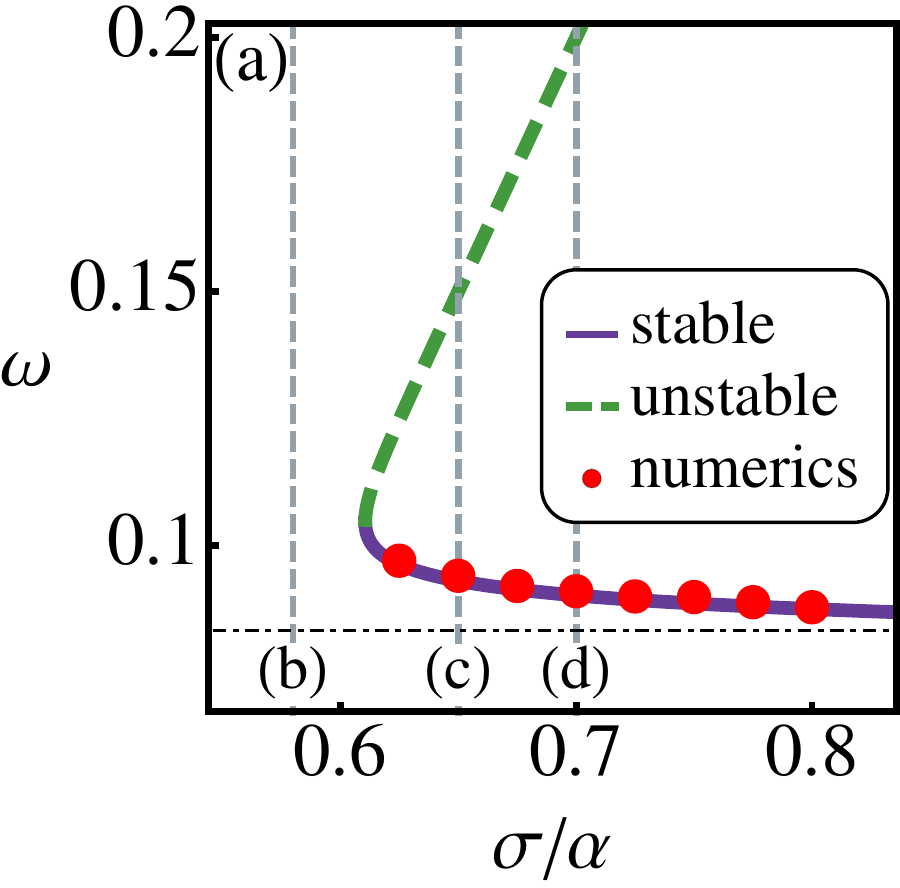}\hspace{1mm}}
  \subfloat{\includegraphics[width=.25\textwidth]{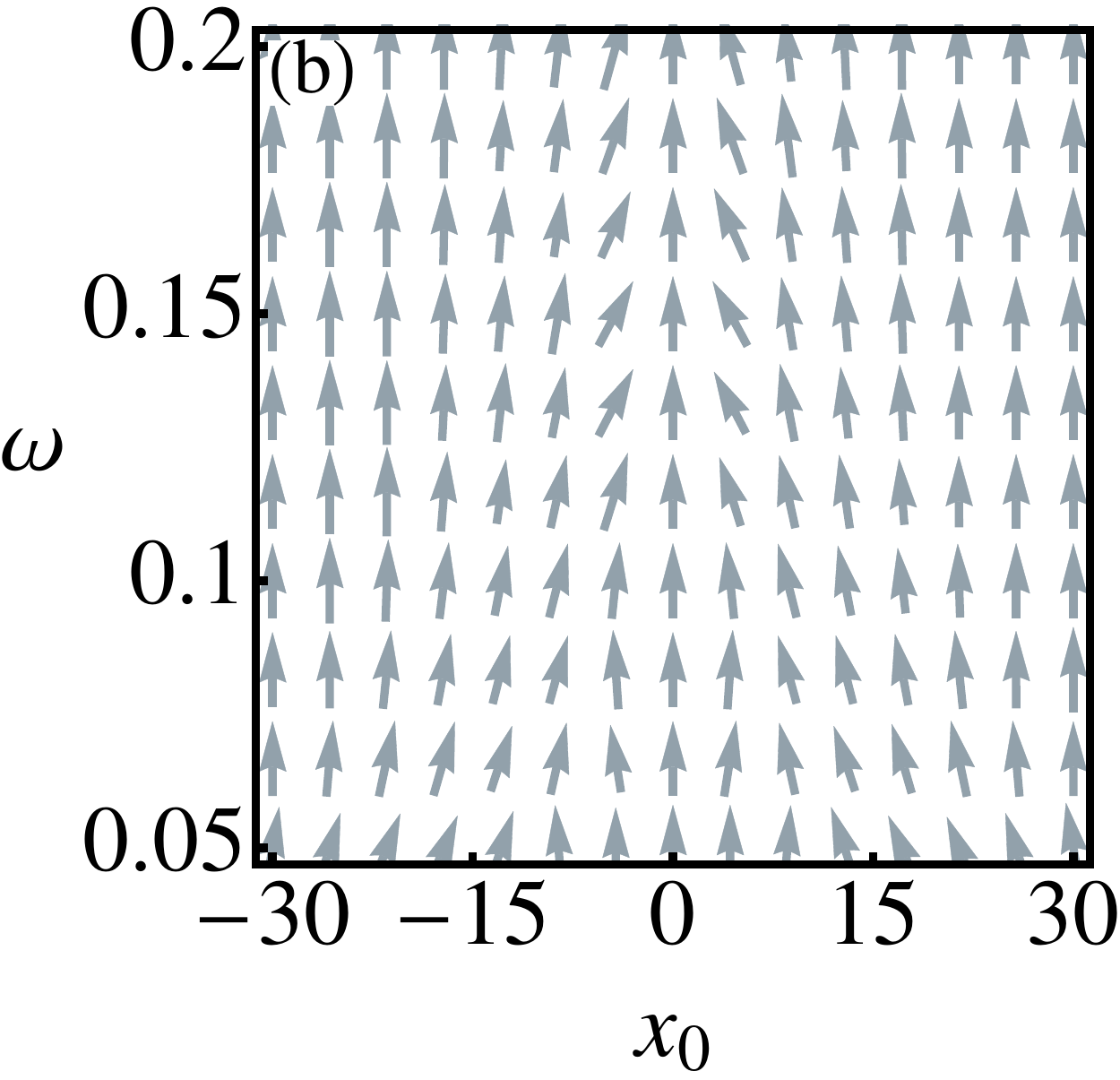}} 
  \\
  \noindent 
  \subfloat{\includegraphics[width=.25\textwidth]{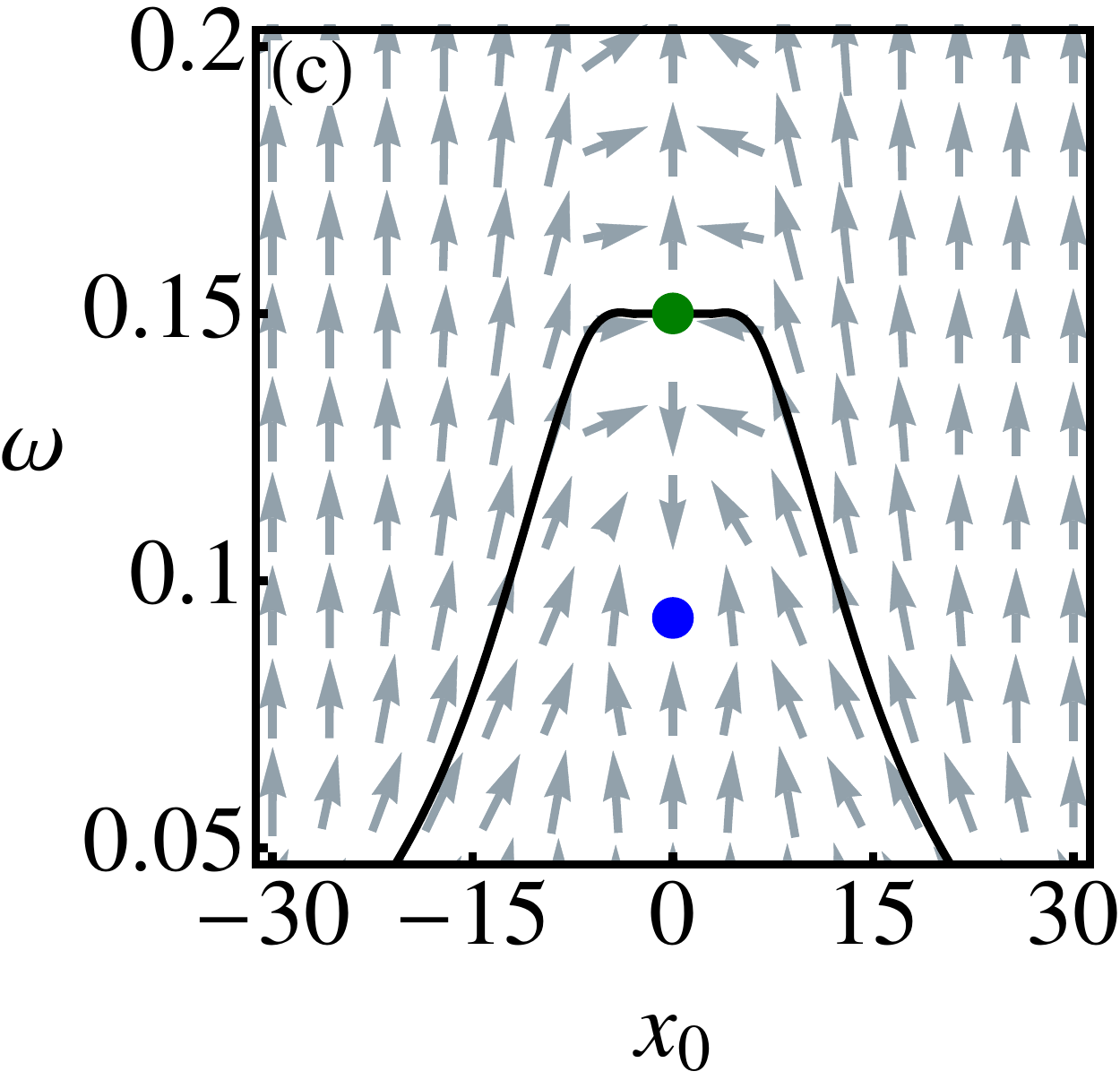}}
  \subfloat{\includegraphics[width=.25\textwidth]{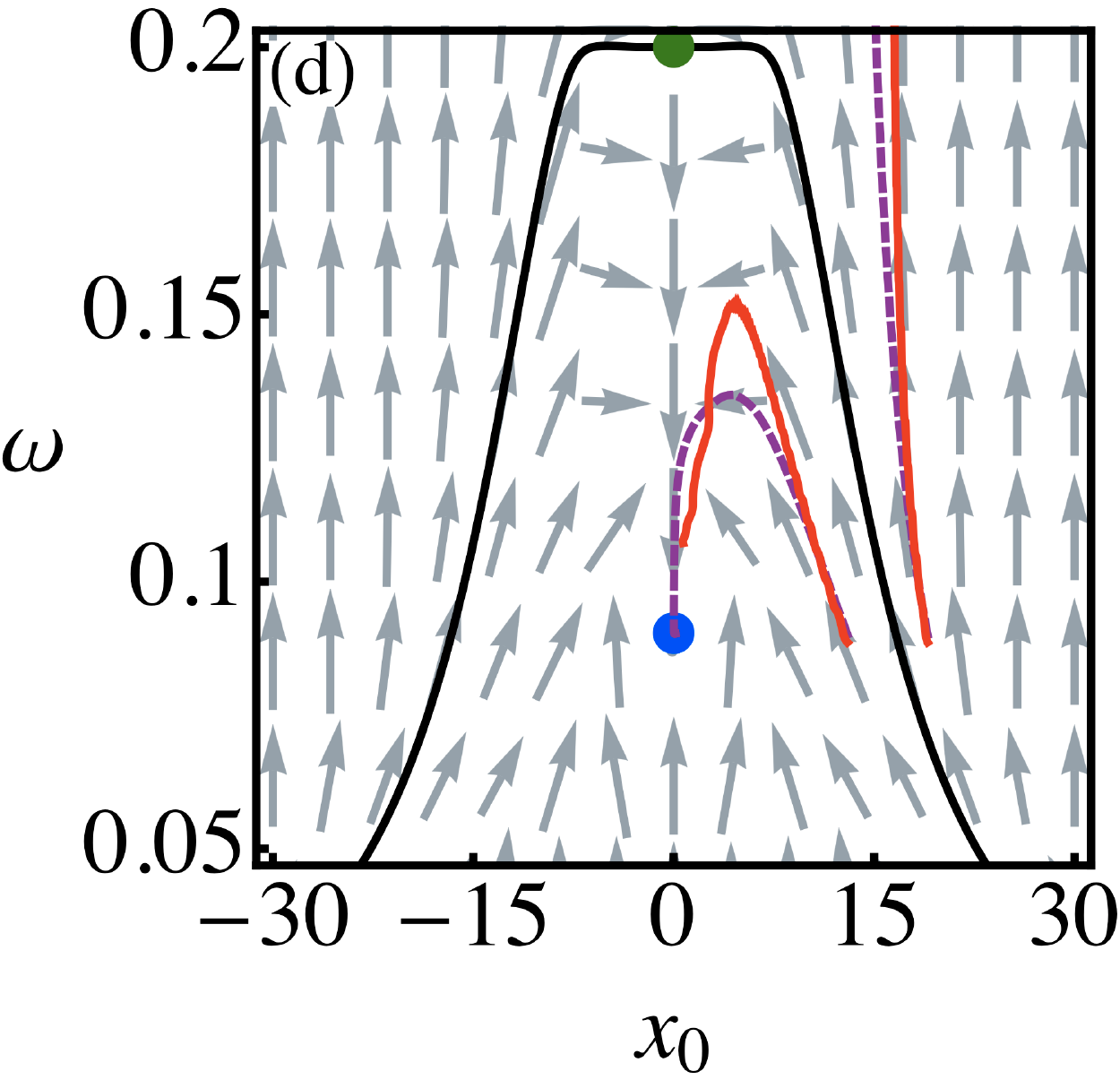}}
  \caption{(a) Dissipative soliton relation \eqref{eq:stabilitycurve}.
    Horizontal line is $\omega = 1/\rho_\star$.  (b-d) ODE vector
    fields corresponding to equations \eqref{eq:5}, \eqref{eq:6} as
    $\sigma$ varies (b) just before the saddle-node bifurcation (c),
    just after and (d) far past bifurcation. The upper/lower dot
    corresponds to the unstable/stable fixed point.  The solid black
    curve encloses the basin of attraction.  Parameters are
    $\rho_\star = 12$, $h_0 = 0.5$, and $\alpha = 0.01$. (d) includes
    trajectories from ODE theory (dashed) and micromagnetics (solid).}
  \label{fig:bifurcationplots}
\end{figure}

We now consider the effects of damping and STT.  A NC-STO consists of
two magnetic layers, one that is assumed fixed and acts as a spin
polarizer of the driving current.  The other layer is dynamic,
resulting from the solution of eq.~\eqref{eq:nondimLL}.  When the
polarizing layer is $\mathbf{z}$, the perturbation $\mathbf{p}$ takes
the form \cite{hoefer2010theory}
\begin{equation}
  p_\Theta = - \alpha \omega \sin \Theta_0 + \sigma V(\rho_\star -
  \rho) \frac{\sin \Theta_0}{1 + \nu \cos \Theta_0}, \quad p_\Phi =
  0, \label{eq:stoperturb}   
\end{equation}
where $\alpha$ is the damping coefficient, $\nu =
(\lambda_{\mathrm{st}}^2 - 1)/(\lambda_{\mathrm{st}}^2 + 1)$,
$\lambda_{\mathrm{st}} \ge 1$ is the spin torque asymmetry,
$\rho_\star$ is the nanocontact radius, and $V$ is a localized
function.  In the following analysis we take $V$ to be the Heaviside
step function thus defining the region of spin polarized current flow
as a disk with radius $\rho_\star$. The STT coefficient $\sigma=
I/I_0$ is proportional to the applied, dc current $I$ with
nondimensionalization $I_0= (\lambda_{\rm
  st}^2+1)M_{\mathrm{s}}^2e\mu_0\pi{\rho_\star}^2\delta/(\hbar
\epsilon \lambda_{\rm st}^2)$ where $\epsilon$ is the spin-torque
polarization, $e$ is the electron charge, and $\hbar$ is the modified
Planck's constant.  For simplicity, we take $\lambda_{\mathrm{st}} =
1$, i.e., no asymmetry. Both $\alpha$ and $\sigma$ are assumed small,
but as a balance must be maintained for the sustenance of a
dissipative soliton \cite{hoefer2010theory}, they are of the same
order.  Substituting this perturbation into
\eqref{eq:mod_w_smallw}--\eqref{eq:mod_x_smallw}, we arrive at a
system of three ordinary differential equations (ODEs). Since
rotational symmetry is not broken for a circular nanocontact, we are
free to rotate the plane and thereby eliminate one of the two
equations for the center.  The modulation system is thus
\begin{align}
  \label{eq:5}
  \frac{d \omega}{dt} = &\alpha \omega^2 ( \omega + h_0) \\
  \nonumber
  &- \frac{\sigma \omega^3}{4 \pi} \int_{|\mathbf{x}|<\rho^*}
  \sech^2(|\mathbf{x}-\mathbf{x}_0|- 1/\omega)\,
  \mathrm{d}\mathbf{x} \\ 
  \label{eq:6}
  \frac{d x_0}{dt} = &- \frac{\sigma \omega^3}{2 \pi}
  \int_{|\mathbf{x}|<\rho^*} \sech^2(|\mathbf{x}-\mathbf{x}_0|-
  1/\omega) \frac{x - x_0}{|\mathbf{x}-\mathbf{x}_0|} \,
  \mathrm{d}\mathbf{x} .
\end{align}
Note that when $h_0 = 0$ and $\sigma = 0$, the remaining ODE $\omega'
= \alpha \omega^3$ agrees with the result in
Ref. \onlinecite{baryakhtar_relaxation_1986} and the more general
result for solitons of non-trivial topological charge in
Ref. \onlinecite{baryakhtar_relaxation_1997}. Equations \eqref{eq:5}
and \eqref{eq:6} do not depend upon the slowly varying phase $\Phi_0$
so that the inclusion of long range magnetostatic effects will lead to
the same frequency shift given in eq.~\eqref{eq:3}, decoupling from
the ODEs \eqref{eq:5} and \eqref{eq:6}.  The fixed points of this
system correspond to steady state conditions where there is a balance
between uniform damping and localized spin torque, i.e., a dissipative
droplet soliton.  A fixed point at $(\omega,x_0) = (\omega_\star,0)$
leads to a relationship between the sustaining current and precession
frequency
\begin{equation}
  \frac{\sigma}{\alpha} = \frac{2  (\omega_{\star} +
    h_{0})}{1 + \omega_\star \log[
    \text{sech}(\rho_\star - \frac{1}{\omega_\star})/2 ]+ \rho_\star
    \tanh( \rho_\star- \frac{1}{\omega_\star})} .
  \label{eq:stabilitycurve}
\end{equation}
Linearizing about the fixed point, we find the eigenvalues
\begin{align}
  \label{eq:7}
  \lambda_1 &= \tfrac{1}{2} \omega_\star [\sigma
  \tanh(\rho_\star - 1/\omega_\star)\\
  \nonumber
  & \qquad \quad +\sigma -\rho_\star
  \sigma\, \text{sech}^2(\rho_\star - 1/\omega_\star)  -2 \alpha h_0], \\
  \lambda_2 &= -\tfrac{1}{2}
  \rho_\star\sigma \omega_\star \,\text{sech}^2(\rho_\star -
  1/\omega_\star) .
\end{align}
For physical parameters, $\lambda_2$ is always negative, however
$\lambda_1$ can change sign as $\omega_\star$ is varied and hence the
stability of the fixed point can change.  This family of fixed points
arises from a saddle-node bifurcation occurring as the current is
increased through the minimum sustaining current
(Fig. \ref{fig:bifurcationplots}(a-d)).  The lower, stable branch of
this saddle node bifurcation is the dissipative soliton.  Figure
\ref{fig:bifurcationplots}(a) shows that the frequency changes little
as the sustaining current is increased from its minimum, stable value.
To a good approximation, the frequency is $\omega_\star =
1/\rho_\star$, as illustrated by the horizontal line in
Fig.~\ref{fig:bifurcationplots}(a). Expanding
\eqref{eq:stabilitycurve} for $\omega_\star$ close to $1/\rho_\star$,
on the stable branch, the fixed point relation can be simplified to
\begin{equation}
  \label{eq:4}
  \frac{\sigma}{\alpha} \sim \frac{2 \left(h_0+\omega_\star
    \right)}{1 + \rho_\star(\rho_\star-1/\omega_\star)} , \quad |\rho_\star -
  1/\omega_\star| \ll 1 . 
\end{equation}
While this relation gives good agreement in the vicinity of the stable
fixed points, it does not predict the minimum sustaining current.
Evaluating $\lambda_1$ for $\sigma$ given by \eqref{eq:4} and
expanding for $0 < \rho_\star - 1/\omega_\star \ll 1$, the eigenvalue
$\lambda_1$ in \eqref{eq:7} is found to be negative, hence the branch
of \eqref{eq:stabilitycurve} nearest to $\omega_\star = 1/\rho_\star$
is indeed stable.  The stable branch is further verified by
micromagnetic simulations (Appendix~\ref{sec:num-methods}) shown in
Fig.~\ref{fig:bifurcationplots}(a).

Interestingly, the dissipative soliton is not a global attractor. The
saddle point's stable manifold (solid curve in
Fig.~\ref{fig:bifurcationplots}(c-d)) denotes the upper boundary in
phase space of the basin of attraction for the dissipative soliton.  A
droplet with frequency $\omega$ and position $x_0$ lying within the
basin of attraction will generally increase in frequency, move toward
the nanocontact center, then decrease in frequency to $\omega_\star$,
converging to the dissipative soliton fixed point.  If an initial
droplet lies outside the basin of attraction, the frequency will
increase, causing the droplet to decrease in amplitude.  An analysis
of the small to moderate amplitude regime
\cite{hoefer_propagation_2012} shows that the soliton decays to spin
waves as its frequency approaches the ferromagnetic resonance
frequency $\omega \to 1$.  Figures \ref{fig:bifurcationplots}(b)-(d)
show the vector field of this system before and after the saddle node
bifurcation.  Figure \ref{fig:bifurcationplots}(d) depicts
trajectories (dashed) generated by numerical evolution of
eqs.~\eqref{eq:5} and \eqref{eq:6} with the initial conditions
$(\omega_\star,13)$ and $(\omega_\star,20)$.  The solid curves result from full micromagnetic
simulations with the same initial conditions.  These numerical
experiments show good agreement up to evolution times
$\mathcal{O}\left(\alpha^{-2}\right)$, as expected for this
approximate theory. The resulting discrepancies lead to modulation
theory slightly over-predicting the radius of the basin of attraction
compared to what is observed from micromagnetics.

\begin{figure}
\includegraphics[width=.5\textwidth]{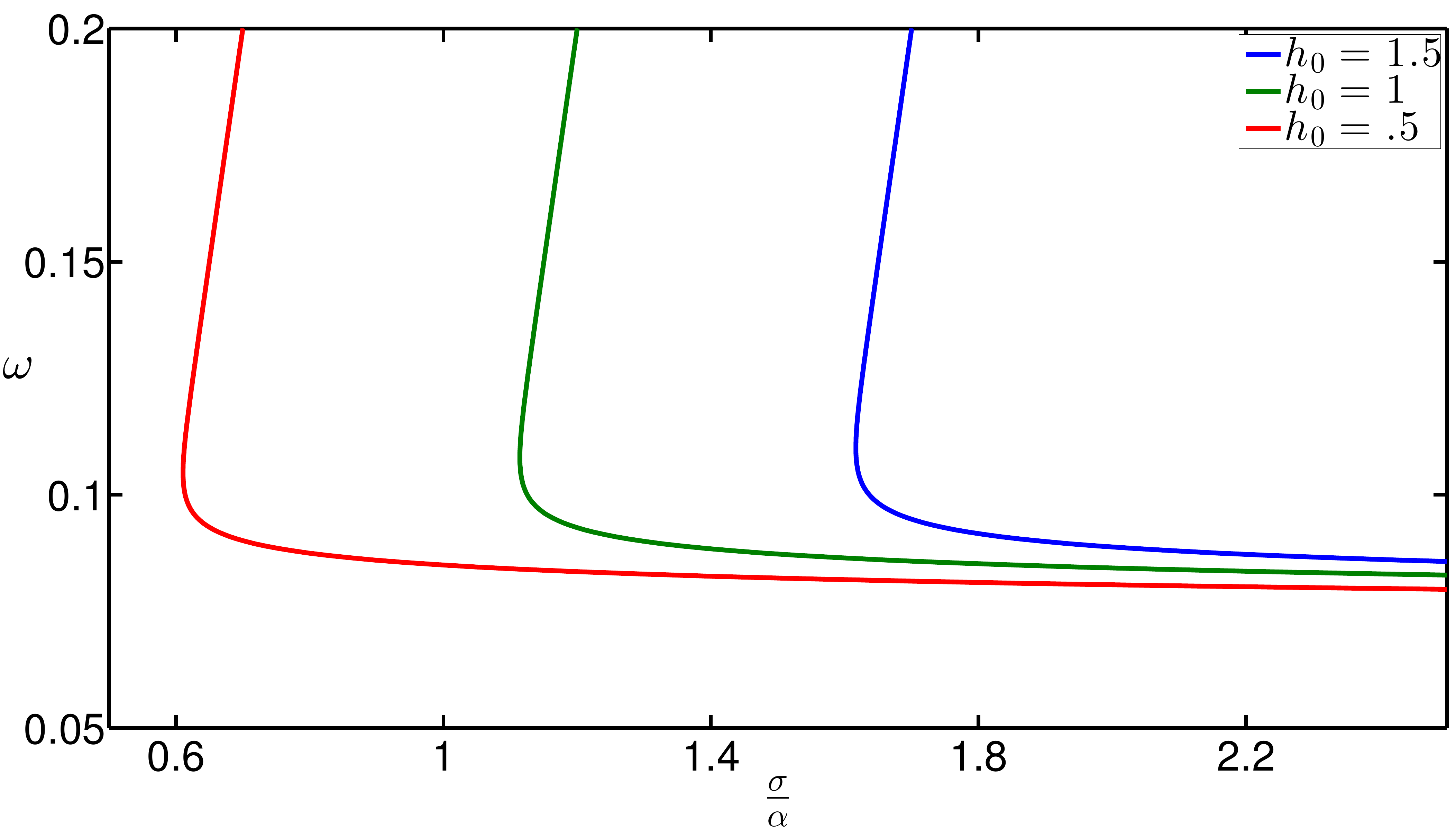}
\caption{Fixed points, both stable and unstable for several values of
  $h_0$.  The primary effect of $h_0$ is to shift these curves of
  fixed points along the $\sigma/\alpha$ axis.}
  \label{fig:h0-stab}
\end{figure}

\begin{figure}
\includegraphics[width=.5\textwidth]{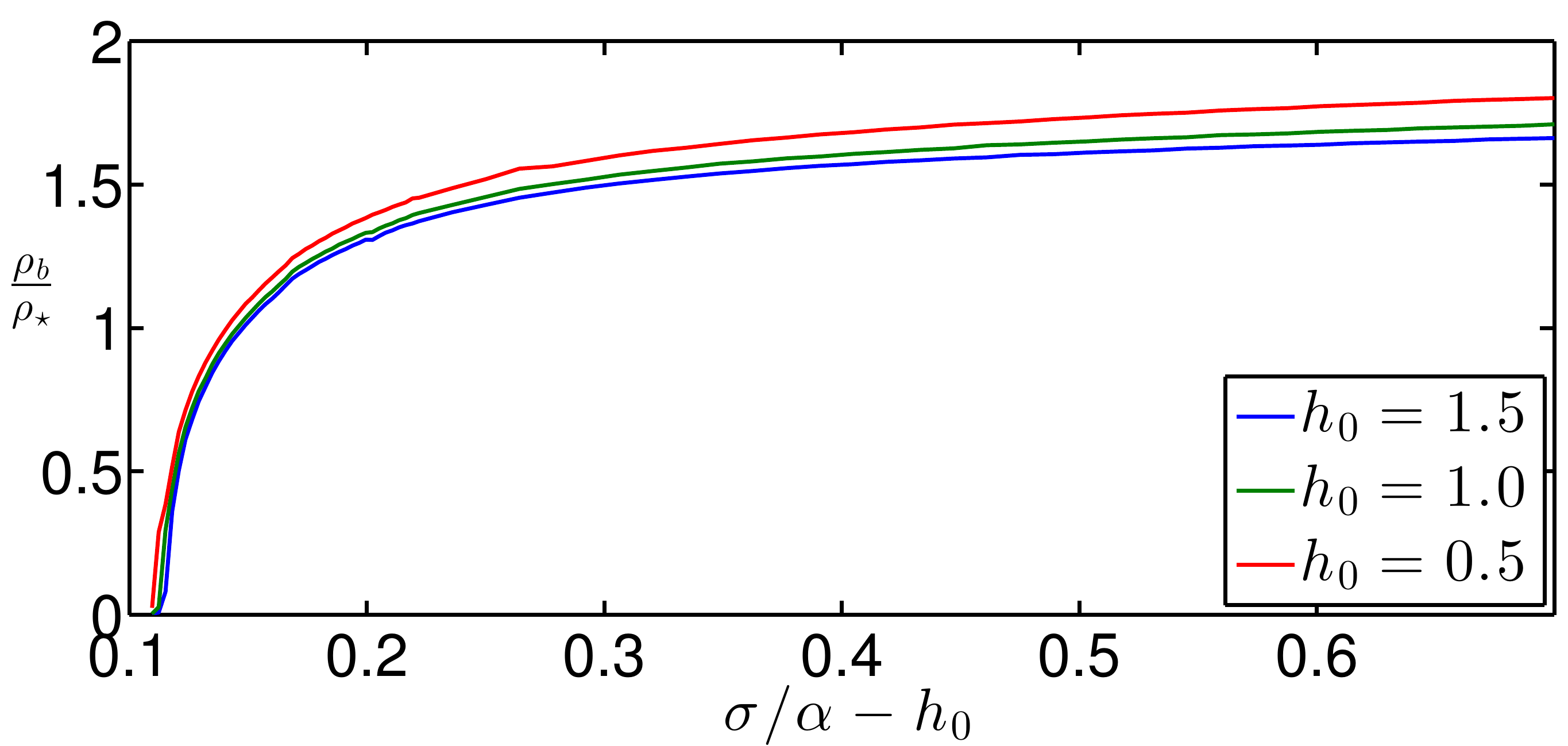}
\caption{Scaled radius of the basin of attraction
  $\rho_{\mathrm{b}}/\rho_\star$ at $\omega = \omega_\star$ with
  $\rho_\star = 12$.  While the center of the
  basin of attraction depends on $h_0$, the width of the basin remains
  essentially unchanged as $h_0$ varies.}
    \label{fig:h0-shift}
\end{figure}

\begin{figure}
  \centering
  \includegraphics[width=8.6cm]{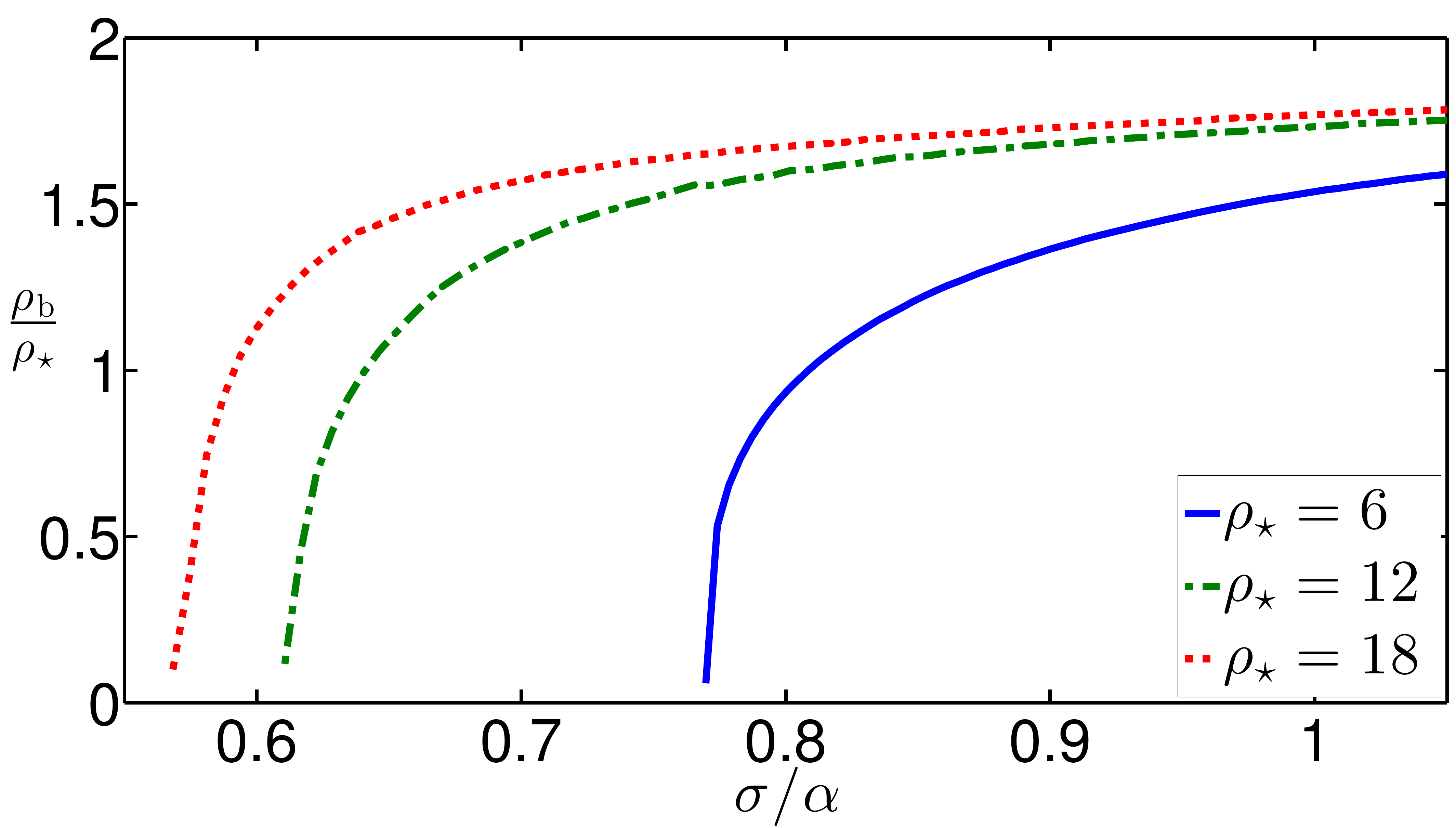}
  \caption{Basin of attraction radius, $\rho_{\rm b}$ at $\omega = \omega_\star$
    scaled by nanocontact radius, $\rho_\star$.  Applied field is $h_0
  = 0.5$. }
  \label{fig:basin_radius}
\end{figure}

The other physical parameters in the fixed point relation
\eqref{eq:stabilitycurve} are $h_0$ and $\rho_\star$.  Based on the analytical form of  eq.~\eqref{eq:stabilitycurve}, 
$h_0$ should shift resulting stability curve. Near the stable branch, the denominator of eq.~\eqref{eq:4} is $\mathcal{O}(1)$ hence an shift of
$\mathcal{O}(h_0)$ is expected. Numerical
experiments with eq.~\eqref{eq:stabilitycurve} suggest that changes in
$h_0$  do essentially serve to shift the minimum sustaining current by a
constant (close to $h_0$). This is apparent from the numerical results
shown in Figs.~\ref{fig:h0-stab} and \ref{fig:h0-shift}.
Figure \ref{fig:basin_radius} depicts the basin of attraction radius
$\rho_{\mathrm{b}}$ (the value of $x_0$ at the edge of the basin of
attraction when $\omega = \omega_\star$) scaled by $\rho_\star$.  As
the current is increased, the basin radius rapidly exceeds
$\frac{3}{2} \rho_\star$ so that a droplet placed well outside the
nanocontact may still experience a restoring force to the nanoncontact
center.

\section{Discussion}
\label{sec:discussion}

We now describe some physical and theoretical implications of the
presented analysis.

The modulation equations
\eqref{eq:mod_w_smallw}-\eqref{eq:mod_x_smallw} are the droplet
soliton analogue of Thiele's equation for magnetic vortices.  They
treat the droplet as a precessing dipole, describing adiabatic changes
in its precessional frequency, phase, and center due to symmetry
breaking perturbations.  This description is valid so long as the
evolution times satisfy $t \ll \epsilon^{-2}$ where $\epsilon$
characterizes the magnitude of the perturbation $\mathbf{p} =
\mathcal{O}(\epsilon)$.  Breaking of the constraint
eq.~\eqref{eq:mod_c_smallw} can lead to an acceleration of the droplet
so that a more general modulation description of propagating droplets
would be required in this case.  Finally, we have neglected coupling
to radiation modes in this analysis which is, for example, important
in the small amplitude regime of NC-STOs
\cite{slonczewski_excitation_1999}.

It is important to point out that the negative frequency shift due to
long range magnetostatic effects is independent of and does not
influence the results pertaining to NC-STO and damping perturbations.
Recalling that $0 < \omega < 1$, stationary droplets in the absence of
long range magnetostatics and applied field are always dynamic.  The
negative frequency shift suggests that a droplet in a sufficiently
thick film could be static, which would correspond to a magnetic
bubble \cite{leeuw_dynamic_1980}.  However, this regime is strictly
outside the validity range of the asymptotic expression \eqref{eq:3}
so it is likely that a non-perturbative analysis is required for
further investigation.

Micromagnetic simulations and experimentally observed low frequency
modulations in Ref.~\onlinecite{mohseni2013spin} suggest the existence
of a restoring force due to the NC-STO.  Our analysis presented here
precisely describes how this restoring force arises, as the
manifestation of a stable fixed point and its basin of attraction.
That the dissipative soliton in the NC-STO is not a global attractor
was observed in micromagnetic simulations previously in the form of
the drift instability \cite{hoefer2010theory}.  As it is known that a
magnetic field gradient can accelerate a stationary droplet
\cite{hoefer_propagation_2012}, we postulate that STT provides a
restoring force that can keep the droplet inside the nanocontact for a
sufficiently small gradient.  These competing effects could also
account for the observed shift of the droplet with respect to the
NC-STO center when Oersted fields are included in the model.  However,
a sufficiently strong field gradient that overcomes the STT restoring
force can lead to expulsion of the droplet, hence a drift instability.
Because a field gradient perturbation does not maintain the constraint
\eqref{eq:mod_c_smallw}, further investigation of this requires the
study of modulated propagating droplets in the presence of an NC-STO.

An additional experimental implication of our results is that of a
restricted regime of droplet soliton excitation.  In NC-STOs, droplets
can be nucleated by a spin wave instability (subcritical Hopf
bifurcation) associated with the uniform state $\mathbf{m} =
\mathbf{z}$ \cite{hoefer2010theory}.  However, the instability may not
generate an excitation that lies within the droplet soliton basin of
attraction.  Furthermore, Fig.~\ref{fig:h0-stab} shows that a large
applied field shifts the minimum sustaining current to higher values.
Since the spin wave instability only weakly depends on the applied
field \cite{slonczewski_excitation_1999,hoefer2010theory}, a
sufficiently large field may shift the stable dissipative soliton
branch above a given applied current so that the droplet is no longer
nucleated in a NC-STO.

\section{Conclusion}
\label{sec:conclusion}

Using singular perturbation theory we have derived modulation
equations for parameters of a droplet soliton under very general
perturbations.  This theory was applied to two such physically
relevant perturbations: 1) higher-order, long range effects of the
magnetostatic field and 2) damping and spin transfer torque forcing in
a nanocontact spin torque oscillator. The key result is that these
long range effects result in a down shift of the overall droplet
frequency. For a NC-STO system, we predict that a droplet shifted from
the nanocontact center can be drawn back by a STT-induced restoring
force.  Sufficiently large shifts cause damping to overwhelm STT
effects so that the droplet soliton is no longer an attractor, hence
decays into spin waves.  For both perturbations investigated we see
good agreement between micromagnetic simulations and the reduced order
models proposed here. The robustness of magnetic droplet solitons to
symmetry breaking perturbations we have demonstrated here suggests
that their initial observation in Ref.~\onlinecite{mohseni2013spin}
represents the beginning of a rich inquiry into novel nonlinear
physics.

\appendix

\section{Approximate Droplet Calculation}
\label{sec:approx-droplet}
Here we offer more detail on the derivation of the small $\omega$
solution to \eqref{eq:droplet}.  For a similar derivation, see
Ref.~\onlinecite{ivanov__1989}.  A uniformly valid approximate
solution to this problem is sought in the limit $0<\w\ll1$. We
begin by introducing a shifted coordinate system $\rho = R +
\frac{A}{\w}$, where $A$ is some constant which will be determined by
solvability conditions. In this coordinate, \eqref{eq:droplet} becomes
\begin{equation}
\label{eq:shifted_droplet}
- \left(\frac{d^2}{d R^2} + \frac{1}{\left(R+\frac{A}{\omega}\right)}
  \frac{d}{dR}  \right) \Theta_0 + \sin \Theta_0 \cos \Theta_0 -
\omega \sin \Theta_0 = 0 
\end{equation}
Expanding \eqref{eq:shifted_droplet} and keeping terms only to leading
order in $\omega$ gives
\begin{equation}
  \label{eq:shifted_droplet_asymptotic}
  - \frac{d^2\Theta_{0} }{d R^2}+ \sin \Theta_0 \cos \Theta_0 +\omega
  \left( -\frac{1}{A} \frac{d \Theta_{0} }{d R} - \sin \Theta_0 \right)
  = \bo{\w^2}. 
\end{equation}
Inserting the asymptotic expansion $\Theta_0 = \Theta_{0,0} + \omega
\Theta_{0,1} +\bo{\omega^2}$ into
\eqref{eq:shifted_droplet_asymptotic} and equating like terms at each
order in $\omega$ we obtain
\begin{align}
  \bo{1}:&- \frac{d^2\Theta_{0,0}}{d R^2} + \sin \Theta_{0,0} \cos
  \Theta_{0,0} = 0 \label{eq:small_omega_O1} \phantom{\bigg|}\\
  \bo{\w}:&- \frac{d^2\Theta_{0,1}}{d R^2} + \cos(2\Theta_{0,0})
  \Theta_{0,1} =\frac{1}{A} \frac{d\Theta_{0,0}}{d R} + \sin
  \Theta_{0,0}. \label{eq:small_omega_Ow}
\end{align}
It is readily verified that the solution to \eqref{eq:small_omega_O1}
is $\Theta_{0,0} = \cos^{-1}\left(\tanh(R +R_0)\right)$ where $R_0$ is
some arbitrary constant. For simplicity, we choose $R_0 = 0$ since it
is not restricted unless we seek a higher order solution.  Taking $L =
-\frac{d^2}{d R^2 } + \cos(2\Theta_{0,0}) $, equation
\eqref{eq:small_omega_Ow} is of the form $L\psi = f$. In this case,
$L$ is a Schr\"{o}dinger operator and hence self-adjoint with kernel
spanned by $\mathrm{sech}(R)$. Solvability then requires that the
right hand side of eq.~\eqref{eq:small_omega_Ow}
$$\frac{1}{A} \frac{d\Theta_{0,0}}{d R}  + \sin \Theta_{0,0}  =
\left(1 - \frac{1}{A} \right) \sech(R),$$ is orthogonal to the kernel
of $L$.  Thus $ \left(1 - \frac{1}{A} \right) \sech(R)$ will be a
nontrivial element of the kernel of $L$ unless $A\equiv 1$. Further,
this choice of $A$ means the equation at $\bo{\omega}$ is trivially
satisfied by taking $\Theta_{0,1} \equiv 0$. Substituting back to the
$\rho$ coordinate system, we obtain the leading order solution
\begin{equation}
  \Theta_0 =  \cos^{-1}\left(\tanh\left(\rho-\frac{1}{\w}\right)\right)
  + \bo{\omega^2}. \label{eq:approximate_droplet}
\end{equation} 
This solution is expected to be valid in the regime when $R$ is
\bo{1}, that is when $\rho$ is of the same order as $1/\omega$. The
residual of eq.~\eqref{eq:droplet} with the approximate solution
\eqref{eq:approximate_droplet} is
\begin{equation*}
  \frac{(1-\rho\w) \sech(\rho - \frac{1}{\omega})}{\rho} .
\end{equation*}
Examination of this residual shows that the approximate solution
\eqref{eq:approximate_droplet} is in fact uniformly valid for all
$\rho$ and introduces deviations at $\bo{\omega^2}$.

% Expanding this approximate solution for the droplet around $\rho  =0$, we observe
% $\Theta_0(0) = \pi$ \textit{to all orders} in $\omega$.  Therefore,
% the approximate droplet is exponentially close to being fully reversed
% at its core.

\section{Modulation Equations Derivation}
\label{sec:mod-eqns}
For this derivation, we rescale the perturbation with a small
parameter $\epsilon$, $\mathbf{p} \to \epsilon \mathbf{p}$, $0 <
\epsilon \ll 1$ and introduce the ``slow'' time $T = \epsilon t$.
The modulated droplet takes the asymptotic form
\begin{align*}
  \Theta(\mathbf{x},t) & =
  \Theta_0(\mathbf{x}+\mathbf{x}_0(T);\omega(T)) + \epsilon
  \Theta_1(\mathbf{x},t,T)+ \ldots \\
  \Phi(\mathbf{x},t) & = \Phi_0(T) + h_0 t + \int_0^t \omega(\epsilon
  t') dt' + \epsilon \frac{\Phi_1(\mathbf{x},t,T)}{\sin(\Theta_0)} +
  \ldots ,
\end{align*}
where $\Theta_0$ and $\Phi_0 + h_0 t + \int_0^t \omega(\epsilon t')
dt'$ represent the conservative droplet with slowly varying
parameters.  Introducing these into the perturbed Landau-Lifshitz
equation, at $\bo{\epsilon}$ we obtain $\psi_t = L \psi +f$.  where
\begin{align}
  L_\Phi &=- \lap + (-\grad \Theta_0 \cdot \grad \Theta_0
  +\cos^2(\Theta_0) - \omega(T)\cos(\Theta_0))  \label{eq:Lphi}\\
  L_\Theta &= -\lap + \cos(2 \Theta_0)- \omega(T)
  \cos(\Theta_0) \label{eq:Lth}\\
  \psi & = \begin{pmatrix} \Theta_1 \\
    \Phi_1 \end{pmatrix} \label{eq:psivec} \\
  L& = \begin{pmatrix} 0 & L_\Phi \\ -L_\Theta &
    0 \end{pmatrix} \label{eq:L}  \mbox{ and } \\
  f & = \begin{pmatrix}p_\Theta - \Theta_{0,\omega} \frac{d\omega}{dT}
    - \grad \Theta_{0} \cdot \frac{d x_0}{d T}) \\ p_\Phi
    -\sin(\Theta_0)\frac{d \Phi_0}{dT} \end{pmatrix} \label{eq:frhs}
\end{align}
Since algebraic growth of either $\Theta_1$ or $\Phi_1$ would lead to
secularity, we apply the condition that $f$ is orthogonal to the
generalized null space of the adjoint of $L$ ($L^\dagger$), denoted by
$N(L^\dagger)$ \cite{weinstein1985modulational}.  By differentiating
the leading order problem with respect to each of the soliton
parameters, $\mathbf{x}_0 = (x_1,x_2)$, $\omega$, $\Phi_0$, and
$\mathbf{V}= (V_1,V_2)$ (we
temporarily allow for moving droplets with velocity $\mathbf{V}$), and
evaluating at the conservative, stationary droplet we obtain
\begin{equation}
  \begin{split}
    N(L^\dagger) = \mbox{span}\Bigg\{ &  \begin{pmatrix} 0 \\
        {\frac{\partial \Theta_0}{\partial x_i}
        }\end{pmatrix},\begin{pmatrix}  \sin(\Theta_0)  \\
        0 \end{pmatrix} ,\\
      &\begin{pmatrix} 0 \\
        \frac{ \partial\Theta}{\partial \omega} \end{pmatrix}
      ,\begin{pmatrix} \sin(\Theta_0) \frac{\partial \Phi}{\partial V_i}
        \\ 0 \end{pmatrix}  \Bigg\}, \quad i=1,2.
  \end{split}
\end{equation}
The vector $(\frac{\partial \Phi}{\partial V_1},\frac{\partial
  \Phi}{\partial V_2})$ is determined according to $(\frac{\partial
  \Phi}{\partial V_1},\frac{\partial \Phi}{\partial V_2})=\tilde{\Phi}
( \cos(\varphi),\sin(\varphi))$, where $\tilde{\Phi}$ satisfies the
boundary value problem
\begin{equation}
  \label{eq:phiv} \hspace{-0.5cm}
  \begin{cases}
    - \left( \frac{\partial^2}{\partial \rho^2} + \frac{1}{\rho}
      \frac{\partial}{\partial \rho} - \frac{1}{\rho^2} \right)
    \tilde{\Phi} \\
    \qquad + \Big[ -\left(\frac{\partial}{\partial
        \rho}\Theta_{0}\right)^2 +\cos^2(\Theta_0) - \omega
    \cos(\Theta_0)\Big]\tilde{\Phi}= -\frac{\partial}{\partial
      \rho}\Theta_{0} \vspace{2mm}\\
    \ds \hspace{1.5cm} \lim_{\rho\rightarrow 0} \tilde{\Phi} \mbox{ is
      finite}, \hspace{2.79cm} \lim_{\rho \rightarrow \infty}
    \tilde{\Phi} = 0.
  \end{cases}
\end{equation}
Applying the solvability condition that $f$ is orthogonal to
$N(L^\dagger)$ gives
\begin{align}
  \frac{d \omega}{ dT} &= \frac{1}{
    \frac{\partial\mathcal{N}}{\partial \omega } } \int_{\mathbb{R}^n}
  \sin(\Theta_0)p_\Theta d
  \mathbf{x}  \label{eq:mod_w}\\
  0& = \int_{\mathbb{R}^n} \grad \Theta_0 p_\phi d
  \mathbf{x}   \label{eq:mod_c}\\
  \frac{d\Phi_0}{dT} & = \frac{1}{\frac{\partial\mathcal{N}}{\partial
      \omega } } \int_{\mathbb{R}^n} \frac{\partial\Theta_0}{\partial
    \omega } p_\phi d
  \mathbf{x}\label{eq:mod_phi} \\
  \frac{d \mathbf{x}_0}{d T} & = \frac{1}{\pi
    \int_0^\infty\tilde{\Phi}\:\frac{\partial\Theta}{\partial\rho}\:
    \rho\: d\rho}
  \int_{\mathbb{R}^n} \sin(\Theta_0) \begin{pmatrix} \cos(\varphi) \\
    \sin(\varphi) \end{pmatrix} \tilde{\Phi} p_\Theta d
  \mathbf{x}.\label{eq:mod_x}
\end{align}
where $\mathcal{N} = \int_{\mathbb{R}^2} (1 - \cos(\theta) )
d\mathbf{x}$ is the total spin. Substituting in the small $\omega$
solution gives equations
\eqref{eq:mod_w_smallw}-\eqref{eq:mod_x_smallw}.

\section{Numerical Methods}
\label{sec:num-methods}

Micromagnetic simulations were performed using a
pseudospectral/Fourier discretization in space and a method of lines
in time similar to those presented in previous work
\cite{hoefer2010theory}. The spatial domain was taken to be square,
typically $[-75,75 ] \times [-75,75]$, large enough that the solution
decayed to zero at the boundary. The mesh width was taken small enough
to provide sufficient decay of the fourier coefficients of the
solution, typically $\Delta x = \Delta y = 0.4$. To improve
convergence properties of this method, the region associated with the
nanocontact was smoothed and approximated by a hyper-gaussian
$\exp(-z^8)$, normalized so that the total
current density is the same as for flow in a cylinder with sharp
edges.  For simulations involving the magnetostatic correction, the
nonlocal terms were implemented in fourier space.

The time evolution was conducted with an adaptive explicit Runge-Kutta
method, with normalization at each time step to preserve unit length
of the magnetization vector.  Initial conditions were chosen to be the
approximate conservative droplet at some frequency generally near the
frequency of the fixed point. The precessional frequency of the
droplet was obtained by examining a fixed spatial point near the edge
of the nanocontact, fitting a line to the time dependence of the
in-plane magnetization phase. For a purely precessional mode, the
frequency of the droplet is the slope of this line.  When $\omega$ is changing in time, we utilize the approximate droplet expansion.  For the approximate solution it holds that
$$\omega^2 = \frac{\mathcal{N} }{4 \int_{\mathbb{R}^2} (x-x_0)^2 (1 - \cos(\Theta_0)) d\mathbf{x} }.$$ This relation was used to extract frequencies from the micromagnetic simulations see in Fig. ~\ref{fig:bifurcationplots}.

\bibliographystyle{apsrev4-1}
\end{document}